# Quench, Glow, or Stay Silent: Distance-Controlled Up-Conversion Emission near Metallic Nanowires

M. Ćwierzona [1,2], M. Żebrowski [1], P. Dziarhai [1], J. Niedziółka-Jönsson [3], M. Jönsson-Niedziółka [3], M. Nyk [4], S. Maćkowski [1], D. Piątkowski [1,5] *

[1] Institute of Physics, Faculty of Physics, Astronomy and Informatics, Nicolaus Copernicus University in Toruń, ul. Grudziądzka 5, 87-100 Toruń, Poland
[2] Université Bourgogne Europe, CNRS, Laboratoire Interdisciplinaire Carnot de Bourgogne ICB UMR 6303, 21000 Dijon, France
[3] Institute of Physical Chemistry Polish Academy of Sciences, Kasprzaka 44/52, 01-224 Warsaw, Poland
[4] Institute of Advanced Materials, Faculty of Chemistry, Wroclaw University of Science and Technology, Wybrzeze Wyspianskiego 27, 50-370 Wroclaw, Poland
[5] Institute of Advanced Studies, Nicolaus Copernicus University in Toruń, Poland

** Corresponding author: dapi@umk.pl*

**Abstract**

We demonstrate controlled transitions between competing radiative and nonradiative decay channels in the up-conversion luminescence of $NaYF_4:Er^{3+}/Yb^{3+}$ nanocrystals placed in proximity to metallic nanowires. The nanocrystal-nanowire separation is used as a key control parameter governing the optical response. An essential aspect of our approach is the removal of inhomogeneous residual polymer layers from the nanowires, eliminating any compromise on distance control and reproducibility in emitter-metal hybrid nanostructures. Replacing them with a well-defined polymer spacer yields controlled access to three qualitatively distinct interaction regimes: luminescence quenching, plasmonic enhancement, and effective decoupling. Transitions between these regimes are shown to reflect changes in the dominant energy relaxation pathways: from nonradiative losses in direct contact with the metal, through modification of the local photonic density of states and coupling to plasmon-mediated modes, to behavior characteristic of quasi-isolated emitters. The plasmonic origin of the emission enhancement in the intermediate regime is evidenced by an increase in luminescence intensity accompanied by shortened decay times, as revealed by fluorescence lifetime imaging microscopy of up-conversion nanocrystals. All experiments were performed on single nanostructures, thereby excluding artefacts arising from aggregation effects and strengthening the interpretation of the observed phenomena. The presented approach provides controlled access to plasmon-modified decay channels and offers a basis for the rational design of functional nanophotonic and sensing structures with tailored optical properties.

## 1. Introduction

Luminescent properties of individual emitters continue to attract significant attention due to their fundamental importance for modern spectroscopy and advanced optical technologies [1]. Nanoemitters are widely employed for monitoring physical and biomedical processes, in telecommunication, quantum technologies, and high-resolution imaging [2-5]. The rapid development of nanomaterial synthesis and engineering strategies in recent years has enabled control over emitter morphology and local environment, allowing emission properties to be tailored on demand [6]. In this context, approaches based on controlled interactions between emitters and metallic nanostructures, metasurfaces, or, more broadly, nanophotonic architectures designed to enhance emission, have attracted considerable interest [7,8]. These strategies integrate diverse concepts for tuning and amplifying luminescent responses, providing a natural framework for investigating control over interactions between individual emitters and their photonic environment.

Despite substantial progress in the design of such systems, achieving quantitatively precise and reproducible control at the level of a single emitter and its immediate surroundings still remains a significant experimental challenge.

Metal-enhanced luminescence (MEL) represents one of the most powerful approach for modifying the emission properties of individual nanoemitters [9]. Two main mechanisms responsible for luminescence enhancement in the presence of metallic nanostructures are distinguished in the literature. First, strong electromagnetic field localization in the vicinity of a metallic nanoparticle increases the local energy density of the excitation radiation, resulting in a higher number of absorption events in the emitter per unit time [10]. Second, according to Fermi's golden rule [11], a metallic nanostructure can modify the local photonic density of states, leading to an increased probability of spontaneous emission [12]. The efficiency of both mechanisms strongly depends on the spectral relations within the hybrid system. Enhancement of absorption requires spectral overlap between the emitter excitation band and the plasmon resonance of the metallic nanostructure, whereas radiative enhancement is most effective when the emission band is red-shifted relative to the plasmon resonance [13]. Appropriate selection of the geometry and material of metallic nanostructures, which define their optical properties, therefore enables the design of hybrid architectures with tailored spectroscopic characteristics. In practice, however, the realization of MEL is limited by the difficulty of achieving precise and reproducible control over the spatial separation between the emitter and the metallic nanostructure, the critical parameter governing the balance between emission enhancement and quenching.

At present, MEL finds application in areas that require increased light-matter interaction at the level of individual emitters, such as ultrasensitive analytics, sensing, and high-resolution optical imaging [14-16]. In these applications, metallic nanostructures act as local electromagnetic field enhancers, enabling the detection of extremely weak optical signals. The effectiveness of this approach relies primarily on precise control over emitter-metal interactions. As demonstrated in canonical experiments employing individual emitters and metallic nanoparticles mounted on atomic force microscopy (AFM) tips, the luminescence enhancement factor peaks at separation distances on the order of 10-15 nm [10]. At shorter distances, nonradiative processes begin to dominate, leading to pronounced emission quenching. At larger separations, the enhancement gradually diminishes, reflecting the reduction in effective coupling between the emitter and the metallic nanostructure. For this reason, control over the separation distance in luminescent hybrid systems is essential for distinguishing and systematically exploring the critical interaction regimes of quenching and enhancement.

Precise manipulation of the emitter-metal separation in metal-enhanced luminescence is of particular importance for nanocrystals (NCs) doped with rare-earth (RE) ions, which in recent years have gained increasing relevance in modern photonics. These nanomaterials are distinguished by chemical stability, well-defined spectrally narrow absorption and emission lines, and the ability to generate anti-Stokes emission through up-conversion processes [17]. Literature reports describe the use of silver island films [18], gold nanoparticles [19], silver nanowires [20], as well as gold AFM tips [21] to modify and control their emission properties. Distance-dependent plasmonic enhancement of up-conversion emission has also been demonstrated using dielectric spacers that allow controlled tuning of the emitter-metal separation [22]. Most of these studies demonstrated that luminescence can be modified, and in some cases enhanced, in the vicinity of plasmonic structures. However, a quantitative interpretation of such interactions remains challenging due to the specific nature of RE-doped nanocrystals. Because the oscillator strengths of f–f transitions in RE ions are very low (~$10^{-6}$) [23], a single nanocrystal, with a diameter ranging from a few to several tens of nanometers, typically contains up to hundreds dopant ions distributed throughout its volume [24]. Each ion occupies a distinct spatial position, and thus the measured luminescence signal represents a superposition of responses from multiple emitters located at different distances from the surface of the metallic nanoparticle. Consequently, optimization of the nanocrystal-metal separation requires a holistic management of competing interaction mechanisms occurring throughout the entire nanocrystal volume.

An additional limitation in studies of MEL arises from the nature of metallic nanoparticles synthesized using wet-chemical approaches. During synthesis, their surface is often covered with inhomogeneous residual surfactant or polymer layers, which directly affect the geometry of any hybrid architectures [25]. As a consequence, the control of spatial separation in emitter-metal systems becomes perturbed, leading to somewhat poor reproducibility and hindering quantitative verification of distance-dependent effects.

These challenges are further compounded in ensemble-averaged measurements, where contributions from nanocrystal agglomerates and spatially heterogeneous coupling conditions cannot be readily distinguished. Moreover, plasmonic architectures designed to maximize resonance overlap with the excitation wavelength simultaneously provide only limited access to emission-related resonances, restricting access into the underlying relaxation pathways. As a result, there is limited amount of generally applicable, reproducible, and controllable platforms that enable systematic optimization of distance-dependent MEL and unambiguous discrimination between emission quenching, enhancement, and decoupling. This gap limits the development of quantitative models of metal-enhanced up-conversion luminescence (MEUCL) as well as controlled fabrication of anti-Stokes luminescent hybrid systems.

In this work, we present a reproducible and physically feasible platform that circumvents these limitations and enables systematic investigations of MEUCL. Silver nanowires (NWs), characterized by a broadband plasmonic response, are employed as a source of localized electromagnetic fields and are separated from RE-doped NCs by a polymer film with controlled thickness.

By operating with well-dispersed nanocrystals and explicitly excluding agglomerates, the proposed approach minimizes ensemble averaging and allows precise spatial localization of emitters with respect to the plasmonic nanostructure. Importantly, combining this architecture with spatially resolved microsecond fluorescence lifetime imaging enables direct correlation between emission intensity, relaxation dynamics, and the local plasmonic environment, thereby providing experimental access to distinct interaction regimes, including luminescence quenching, plasmon-enhanced emission, and effective emitter-metal decoupling. The simplicity and scalability of the fabrication procedure make this platform attractive for reproducible studies of plasmon-modified up-conversion processes and for the rational design of anti-Stokes hybrid systems.

## 2. Materials and Methods

A hybrid sample architecture comprising silver NWs, a polymer film with a defined thickness, and NCs was devised and used in the experiments. In this geometry, the controlled thickness of the poly(vinyl alcohol) (PVA) spacer film defines the minimum achievable separation between the up-conversion NCs and the silver NWs, as schematically presented in Fig. 1.

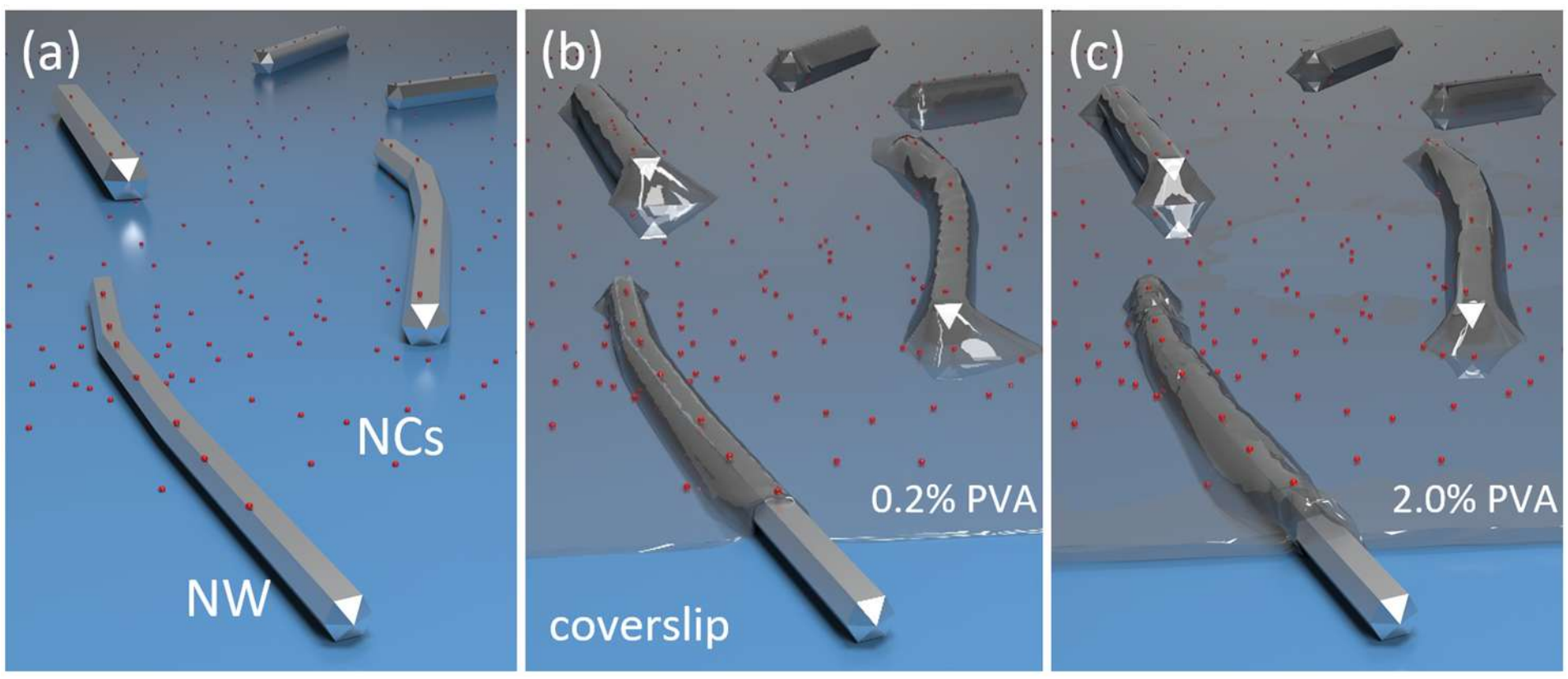

*Fig. 1. Schematic illustration of controlled spatial geometry of up-conversion NCs and silver NWs. Increasing the thickness of the PVA film in the sequence (a)-(c) provides a means to systematically vary their spatial separation.*

The nanowires, serving as the key plasmonic component of the hybrid nanostructure, were synthesized using a wet-chemical polyol process [26]. They exhibit an average diameter of 205±40 nm and lengths reaching from several to several tens of micrometers, as shown in a scanning electron microscopy (SEM) image in Fig. 2a (Neoskop JCM-7000, JEOL). As-synthesized nanowires are coated with a poly(vinylpyrrolidone) (PVP) layer of non-uniform thickness, ranging typically from a few up to about 20 nm. However, in some regions this layer is practically absent, while locally it may even exceed 100 nm, as shown in the electron microscopy images in Fig. 2b (Nova NanoSEM 450, FEI). Silver nanowires are characterized by a broad absorption band extending across the visible and near-infrared spectral ranges [27]. This unique feature makes them particularly suitable for up-conversion-based emitters characterized by large anti-Stokes shifts, for modifying both their excitation (absorption) and emission pathways (Fig. 2c).

In our approach, we first remove the inhomogeneous residual PVP layer, identified as a primary source of structural nonuniformity. For that purpose, we applied a solvent-based washing procedure commonly used to reduce the PVP layer thickness [28], based on alternating washing in acetone and ethanol baths. The cleaned nanowires were subsequently dispersed in water and deposited onto glass substrates (microscope coverslips) by spin-coating 20 µL of the colloid at 2500 rpm for 20 s. Appropriately adjusted colloid concentration enabled the deposition of individual, well-separated NWs on the substrate, with a surface density of approximately 2-3 nanowires per 100 µm$^2$.

To control the separation between the NCs and the silver NWs, thin polymer films were deposited onto the pre-positioned NWs by spin-coating 20 µL of the PVA solution at 2500 rpm for 30 s. Polymer solutions with concentrations of 0.2% and 2.0% were used, yielding films with thicknesses of 20±5 nm (sample S20) and 65±10 nm (sample S65), respectively. The thickness of the PVA film was determined by profilometric step-height measurements performed on a reference scratch for several independently prepared samples, as described in Supplementary Information (SI). For reference, a sample without the spacer was also prepared and labeled S00.

In the final step of sample preparation, a dispersion of $NaYF_4$ nanocrystals doped with 2 mol% $Er^{3+}$ and 20 mol% $Yb^{3+}$ ions was deposited onto the NW substrates by spin-coating 20 µL of the solution at 2500 rpm for 20 s. The NCs were synthesized according to a previously established procedure [29]. Their average diameter is approximately 30 nm, as shown in a transmission electron microscopy (TEM) image in Fig. 3a (Tecnai G$^2$ 20 X-TWIN, FEI), and when dispersed in chloroform they form a stable colloidal suspension. After spin-coating on a glass substrate, the nanocrystals are typically distributed as well-isolated individual objects, showing no significant tendency toward agglomeration, as can be seen in an SEM image in Fig. 3b (Nova NanoSEM 450, FEI) and photoluminescence (PL) image in Fig. 3c. The $Er^{3+}$/$Yb^{3+}$co-doped system represents a canonical and well-characterized up-conversion platform [30]. In this case, $Yb^{3+}$ ions act as sensitizers, efficiently absorbing excitation radiation at 980 nm and subsequently transferring energy nonradiatively to $Er^{3+}$ ions, where sequential excitation to higher-lying energy levels occurs (Fig. S2, SI). The emission spectrum is characterized by two intense anti-Stokes green and red luminescence bands, centered around 540 nm ($^2H_{11/2}$,$^4S_{3/2}\rightarrow{}^4I_{15/2}$) and 650 nm ($^4F_{9/2}\rightarrow{}^4I_{15/2}$), respectively (Fig. 3d).

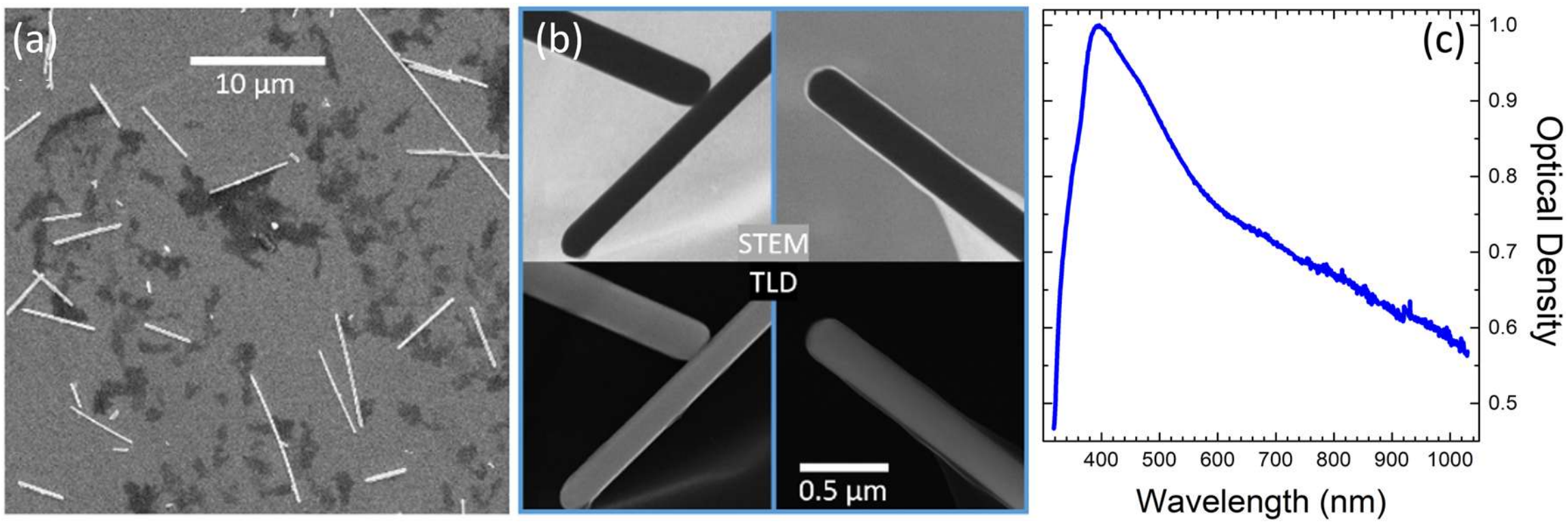

*Fig. 2. (a) SEM image of silver nanowires deposited on a conductive carbon foil. (b) Scanning transmission electron microscopy (STEM) images of nanowires (upper panel) illustrating the surrounding PVP polymer layer, with corresponding images acquired using a through-lens detector (TLD) shown in the lower panel for comparison. (c) Absorbance spectrum of the silver nanowire colloid.*

The experimental setup was based on an inverted optical microscope (Eclipse Ti2, Nikon), equipped with a high-numerical-aperture oil-immersion objective (Apo TIRF 60×, NA=1.49), as schematically shown in Fig. S3 (SI). Nanocrystals were excited using a laser diode (LP980-SF15, Thorlabs) operating at 980 nm, producing a diffraction-limited excitation spot on the sample surface. The luminescence was collected by the same objective and directed by a dichroic mirror (T926spxrxt, Chroma) through a confocal pinhole (d=25 µm) into the detection path. For photoluminescence imaging, the signal passed through a bandpass filter (FB550-40 or FB650-40, Thorlabs) and was detected by a single-photon counting module (SPCM, COUNT-100C, Laser Components), connected to a pulse-counting card (PCI-6236, National Instruments) and synchronized with a piezoelectric sample stage (P-545, Physik Instrumente) operating in a raster-scanning mode. In this configuration, the system enabled acquisition of PL intensity maps or back-scattered (SC) laser light images with an effective spatial resolution (FWHM) of approximately 400 nm. For time-resolved measurements, the SPCM was coupled to a multiscaler card (PMS-400A, Becker & Hickl), allowing registration of luminescence decay transients with a temporal resolution of 1 µs. Alternatively, combining imaging with time-resolved detection enabled pixel-by-pixel mapping of the spatial distribution of luminescence decay times, referred to as microsecond fluorescence lifetime imaging microscopy (µs-FLIM). In addition, the luminescence signal could be directed to a monochromator (Shamrock 500i, Andor Technology) coupled to a CCD camera (iDUS 420, Andor Technology) for collecting the emission spectra.

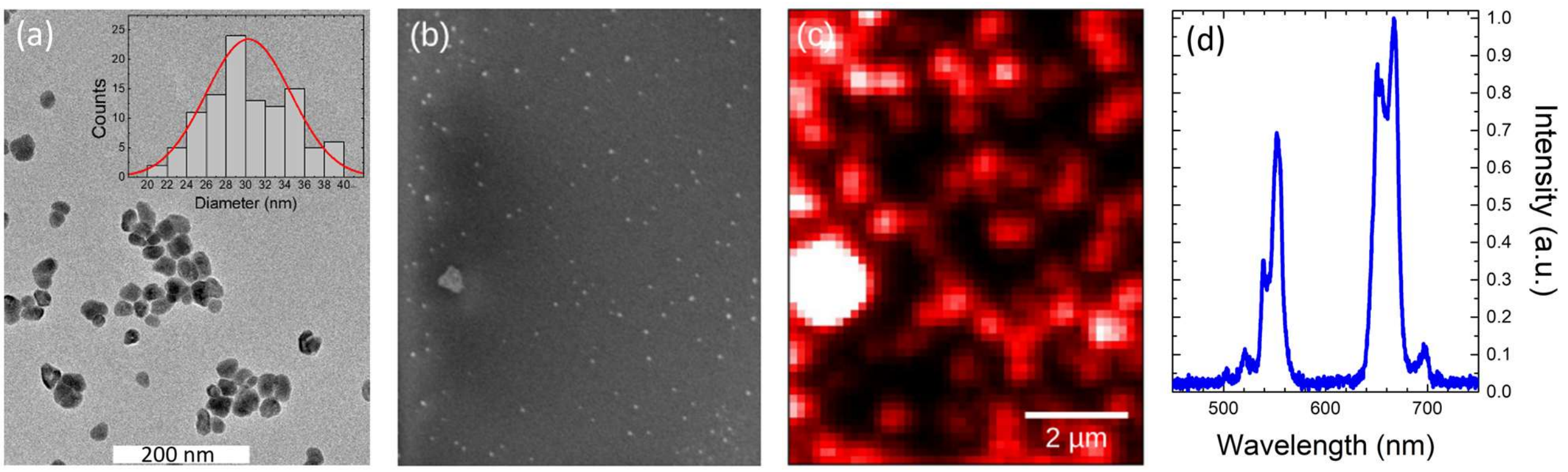

*Fig. 3. (a) TEM image of the $NaYF_4:Er^{3+}/Yb^{3+}$ NCs together with a histogram illustrating the statistical size distribution. (b) SEM image of individual NCs dispersed on a glass substrate (S00) and the corresponding PL image of the same sample area (c). Only minimal agglomeration is observed, which is easily distinguishable in both SEM and PL images. (d) Emission spectrum of a single NC under 980 nm excitation.*

## 3. Results and discussion

Luminescence imaging performed for all three samples under 980 nm excitation at a power of 10 mW reveals clear differences in emission behavior (Fig. 4). Since the green and red emissions of $Er^{3+}$ arise from the same physical processes [31], the emission intensity analysis presented here is primarily focused on the red emission, which provides a higher signal-to-noise ratio. The recorded maps of red up-conversion PL reveal relatively dense, yet still dispersed, distributions of nanocrystals across the sample surface (Fig. 4a-c). High-intensity spots are observed only in rare cases and are attributed to nanocrystal agglomerates. These regions were excluded from further analysis.

For sample S00, where NCs were deposited directly onto the substrate containing the NWs, ensuring the minimal separation distance between the nanoparticles ($d \rightarrow 0$), the PL map reveals pronounced regions of strongly reduced emission intensity (Fig. 4a). These regions appear as elongated, dark shapes observed against the background of relatively uniform NC emission. Correlation of the PL map with the image of back-scattered laser light (Fig. 4d) allows us to assign these regions to the positions of the silver NWs. The SEM image acquired for this area (inset in Fig. 4a) confirms the presence of NCs on the nanowire surface, indicating that the observed quenching of luminescence does not arise from the absence of NCs.

Photoluminescence map recorded for sample S20, with NCs-NW separation of 20 nm, exhibits a clearly different behavior compared to the S00 sample. Instead of dark regions correlated with NW positions, bright stripes with enhanced emission intensity are observed (Fig. 4b). The typical PL intensity reaches approximately 500 kcps and is over ten times higher than that measured in regions located away from the silver nanowires. Correlation between the PL and SC maps indicates that the regions of high emission intensity coincide with the positions of the NWs. Importantly, SEM images obtained for the same sample area do not reveal any substantial increase in nanocrystal density in these regions that could account for the observed effect (SI, Fig. S4).

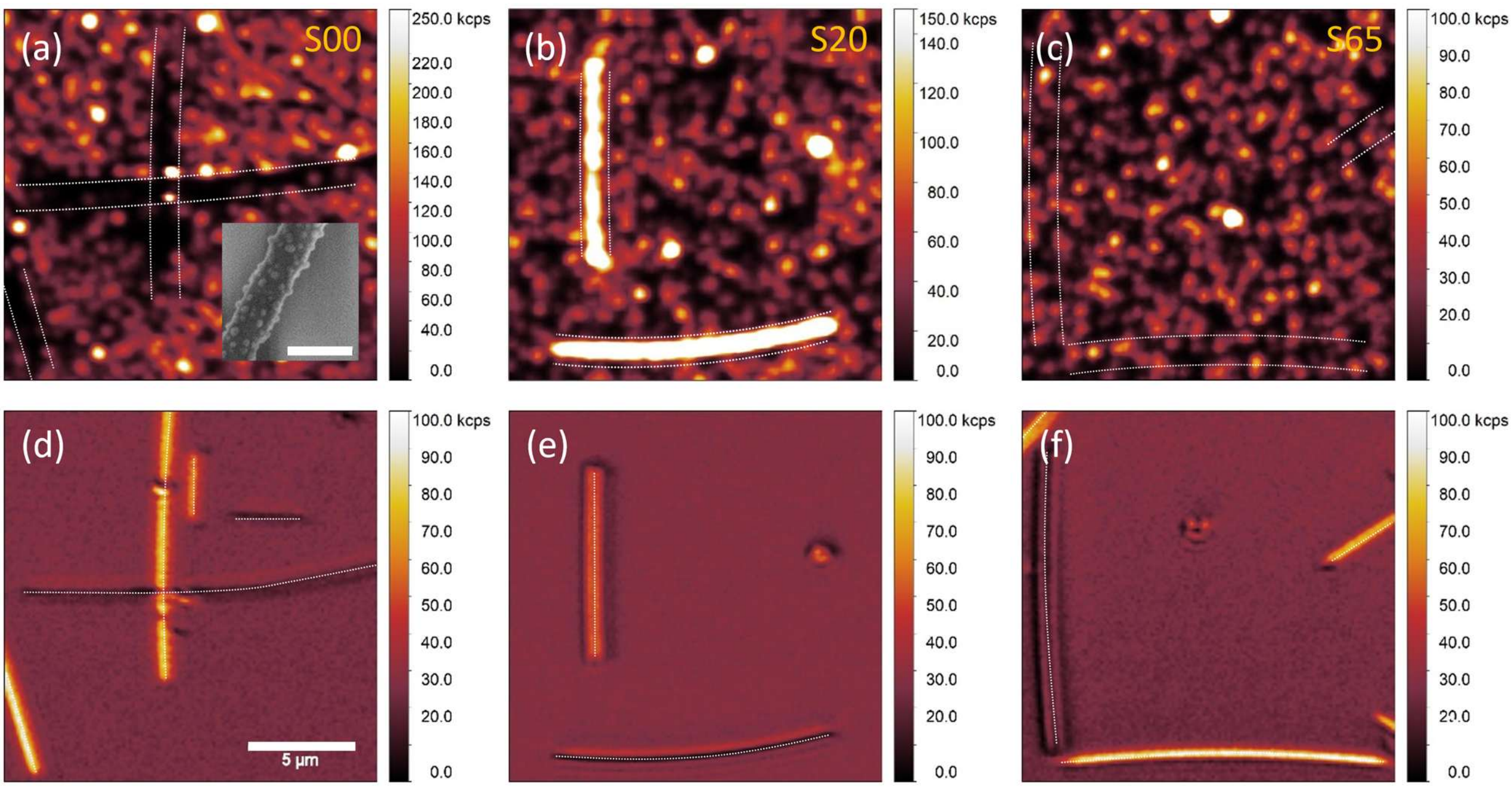

*Fig. 4. Photoluminescence intensity maps of individual up-conversion NCs acquired for samples S00, S20, and S65 (a-c), together with the corresponding maps of elastically back-scattered laser light (d-f), illustrating the spatial distribution of NWs. The white dashed lines indicate the positions of the NWs. The PL maps were obtained by recording the red up-conversion emission of the NCs under 980 nm laser excitation. The inset in panel (a) shows an SEM image of a bare silver nanowire with attached nanocrystals. The scale bar represents 200 nm.*

The luminescence intensity map recorded for sample S65, characterized by the largest separation between the nanocrystals and the nanowires (d=65 nm), is shown in Fig. 4c. The PL image exhibits a random distribution of nanocrystals, without any presence of well-defined regions exhibiting either enhanced or reduced emission intensity. Comparison of the PL map with the SC map (Fig. 4f) reveals no difference in the emission intensity of nanocrystals located close to the nanowires and farther away. The emission intensity distribution remains uniform across the entire investigated area lines.

To further elucidate the origin of the bright stripes observed in sample S20, time-resolved µs-FLIM measurements were performed on the same sample region, providing insight into the spatial distribution of the emission dynamics. The maps obtained for the green and red emissions are shown in Fig. 5a and b, respectively. They show that in regions of enhanced PL intensity the luminescence decay times are approximately 60 µs and 100 µs for the green and red emission, respectively. In contrast, for nanocrystals located away from the metallic structures, the average decay times are approximately 90 µs and 120 µs.

Analysis of the collected data indicates the emergence of three distinct interaction regimes between up-converting nanocrystals and metallic nanostructures, determined by their mutual geometric configuration. These regimes include luminescence quenching, luminescence enhancement, and effective decoupling.

The absence of nanocrystal emission in regions where silver nanowires are present, observed for sample S00, unambiguously indicates strong emission suppression. Indeed, intentional removal of the polymer from the nanowire surface results in extremely small distances between the nanocrystals and the nanowires. Under such conditions, the coupling becomes strong, and energy is efficiently transferred nonradiatively from the NCs to the NWs, where it is dissipated via ohmic losses [32]. A similar quenching mechanism has been observed for molecular emitters and quantum dots located in close proximity to metallic surfaces [10].

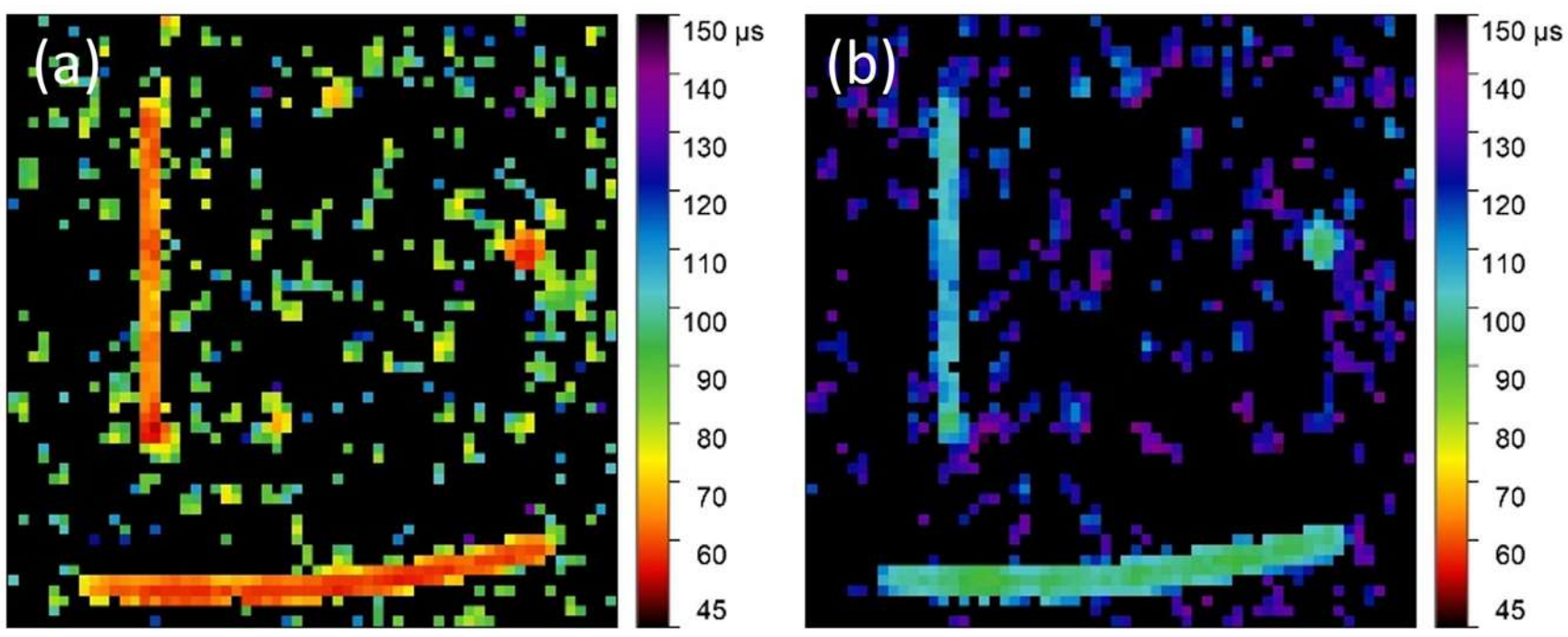

*Fig. 5. µs-FLIM maps of the up-conversion emission recorded for sample S20, acquired from the same sample region as shown in Fig. 4b,e. Panel (a) shows the green emission, while panel (b) corresponds to the red emission of the NCs. Both maps were obtained under 980 nm laser excitation using a pulsed excitation scheme with a pulse duration of 20 µs. The color scale represents the extracted luminescence lifetime values.*

In the case of nanocrystals doped with RE ions, this effect is particularly efficient, primarily due to very small oscillator strengths of radiative f–f transitions, which make these systems susceptible to competing nonradiative relaxation channels.

The presence of a PVA layer with a thickness of 20 nm leads to a qualitative change in the nature of the interactions. The regions of enhanced emission intensity, spatially correlated with the nanowires, indicate that the system enters a regime of enhanced luminescence. A key confirmation of the plasmonic origin of this effect is the shortening of the luminescence decay time observed in the µs-FLIM maps. Although an increase in emission intensity is most commonly attributed to enhanced absorption [20], the simultaneous reduction of the luminescence lifetime observed here indicates an increased decay rate, consistent with a modification of the local photonic density of states near the NWs and the opening of additional emission pathways, including coupling to surface plasmon polariton modes [24]. This interpretation is supported by the fact that the applied spacer thickness falls within the range of distances commonly regarded as optimal for the occurrence of the metal-enhanced luminescence effect [10]. It should be emphasized, however, that in contrast to single dipoles, the up-converting nanocrystals contain a spatially extended population of luminescent centers. The observed enhancement therefore represents an averaged response of the entire ion population, the majority of which is located within distances where modification of the local photonic density of states dominates over possible nonradiative losses, thus leading to a net enhancement of the emission from the NCs.

The absence of any visible correlation between the PL and SC maps observed for sample S65 indicates that for a PVA film thickness of 65 nm, the NCs-NW interaction is suppressed. This separation distance exceeds the range of efficient near-field coupling associated with the plasmonic resonance of the NWs, leading to the disappearance of both absorption enhancement and any significant modification of the local photonic density of states. As a consequence, the NCs behave as quasi-isolated emitters, with their emission properties primarily governed by the surrounding dielectric polymer medium. It is worth emphasizing that the presence of a thick polymer spacer modifies these local conditions, potentially leading to statistical variations in nanocrystal emission intensity unrelated to direct coupling with metallic nanostructures. This is indirectly evidenced by changes in the intensity of nanocrystals located away from the NWs, as noted for samples S00 and S20.

The results presented in Fig. 4 clearly demonstrate that a polymer layer with a controlled thickness can be used to switch between three fundamental interaction regimes in a NW-NCs hybrid nanostructure: strong quenching (S00), metal-enhanced emission (S20), and decoupling (S65). Importantly, all these qualitatively distinct MEUCL regimes can be intentionally and reproducibly achieved within a single, geometrically well-defined platform.

A key advantage of the proposed architecture is its simplicity and high reproducibility. The geometric separation between NCs and NWs can be defined independently of any uncontrolled residual surfactant layers. In contrast to approaches based on random positioning of emitters with respect to plasmonic structures, the presented solution enables systematic and quantitative investigation of distance-dependent effects in photon up-converting systems. This approach, which is not sample-specific, can be straightforwardly adapted to other classes of emitters and plasmonic architectures, offering a universal platform for studies of nanoscale control of optical properties.

## 4. Conclusions

In this work, we present a simple, reproducible model platform for controlled studies of plasmon-modified up-conversion emission in $NaYF_4$ nanocrystals co-doped with $Er^{3+}$ and $Yb^{3+}$ ions. By employing silver nanowires in combination with a polymer spacer layer of controlled thickness, deterministic access to three qualitatively distinct emitter-metal interaction regimes is achieved: emission quenching, metal-enhanced up-conversion luminescence, and effective decoupling of the system. A key element of the proposed architecture is the elimination of uncontrolled PVP layers covering the metallic nanostructures and their replacement with a well-defined spacer, enabling unambiguous interpretation of the observed distance-dependent effects. The plasmonic origin of the emission enhancement is confirmed by a correlated increase in luminescence intensity accompanied by a shortening of the luminescence decay times observed in µs-FLIM imaging. Overall, the proposed approach provides a versatile platform for the rational design of hybrid up-converting systems with controlled emission properties, where appropriate nanoplatform design suppresses detrimental interactions while enabling plasmon-induced emission enhancement.

## Acknowledgments

MĆ and DP were funded by the SONATA BIS project of the National Science Centre, Poland, under project no. 2017/26/E/ST3/00209.

**Supplementary Information for**

# Quench, Glow, or Stay Silent: Distance-Controlled Up-Conversion Emission near Metallic Nanowires

**by**

M. Ćwierzona, M. Żebrowski, P. Dziarhai, J. Niedziółka-Jönsson,
M. Jönsson-Niedziółka, M. Nyk, S. Maćkowski, D. Piątkowski

## Thickness Measurements of PVA Films

The polymer film thickness was evaluated by scratch step-height analysis performed on the sample surface using an optical profilometer (Bruker ContourGT-K0). Scratches penetrating through the polymer layer down to the underlying substrate were intentionally introduced using a syringe tip, and the resulting height profiles across the scratch edges were analyzed. For each polymer concentration, several dozen independent scratches per sample were examined to ensure statistical relevance. Identical measurements were also carried out on the bare glass substrate in order to determine the characteristic reference response of the substrate itself (Fig. S1). The average scratch depth measured on glass was found to be $h_{ref}$ = 20 ± 5 nm and was treated as a systematic contribution related to the scratching and measurement procedure. After accounting for this reference value, the effective thickness of the polymer layer was estimated to be 20±5 nm for the polymer concentration of 0.2%, and 65±10 nm for 2% PVA concentration.

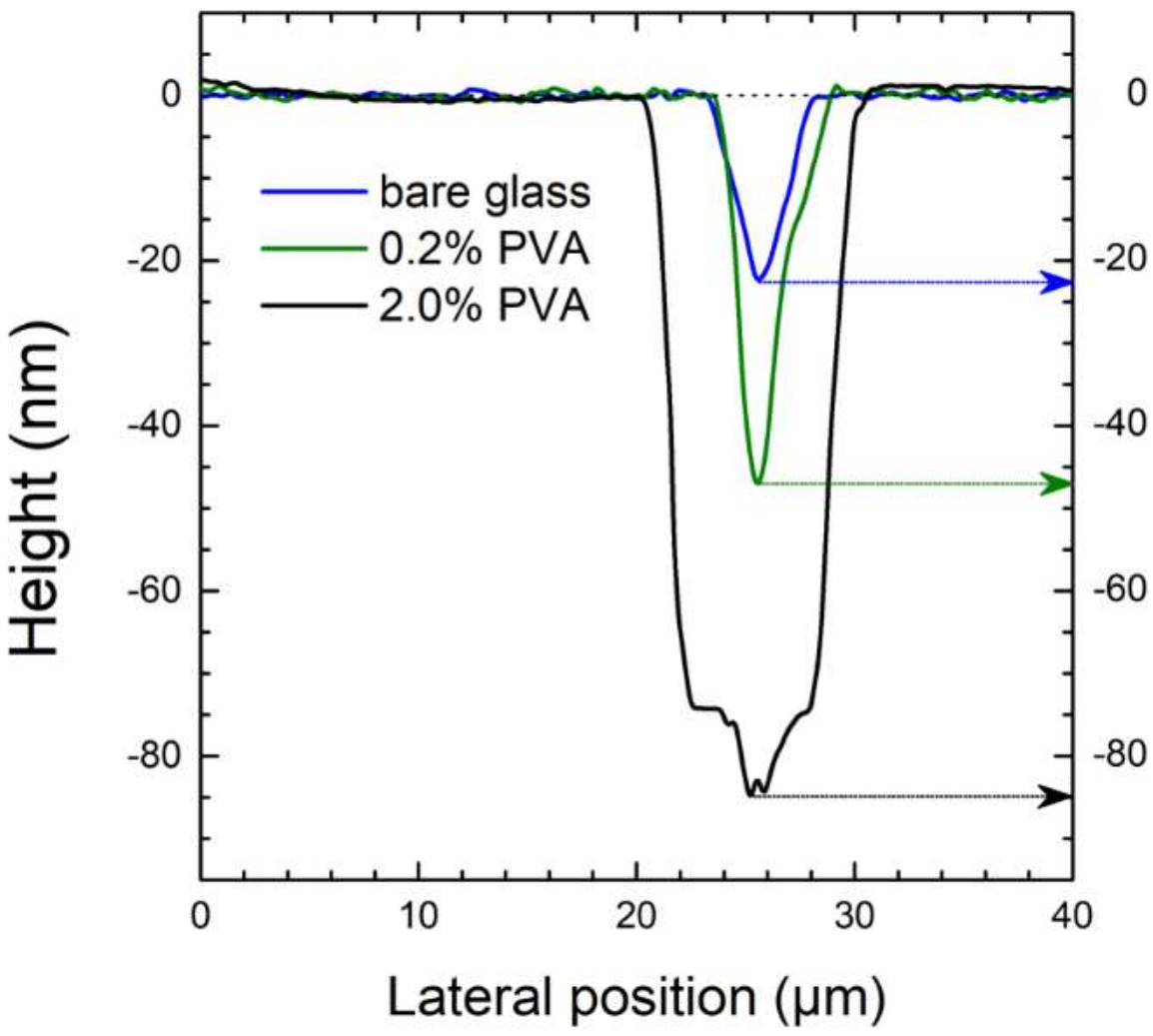

*Fig. S1. Representative profilometric height profiles across scratch cross-sections measured on bare glass (blue) serving as a reference and on PVA films prepared from 0.2% (green) and 2.0% (black) polymer solutions. The shown profiles represent raw scratch depths prior to reference subtraction.*

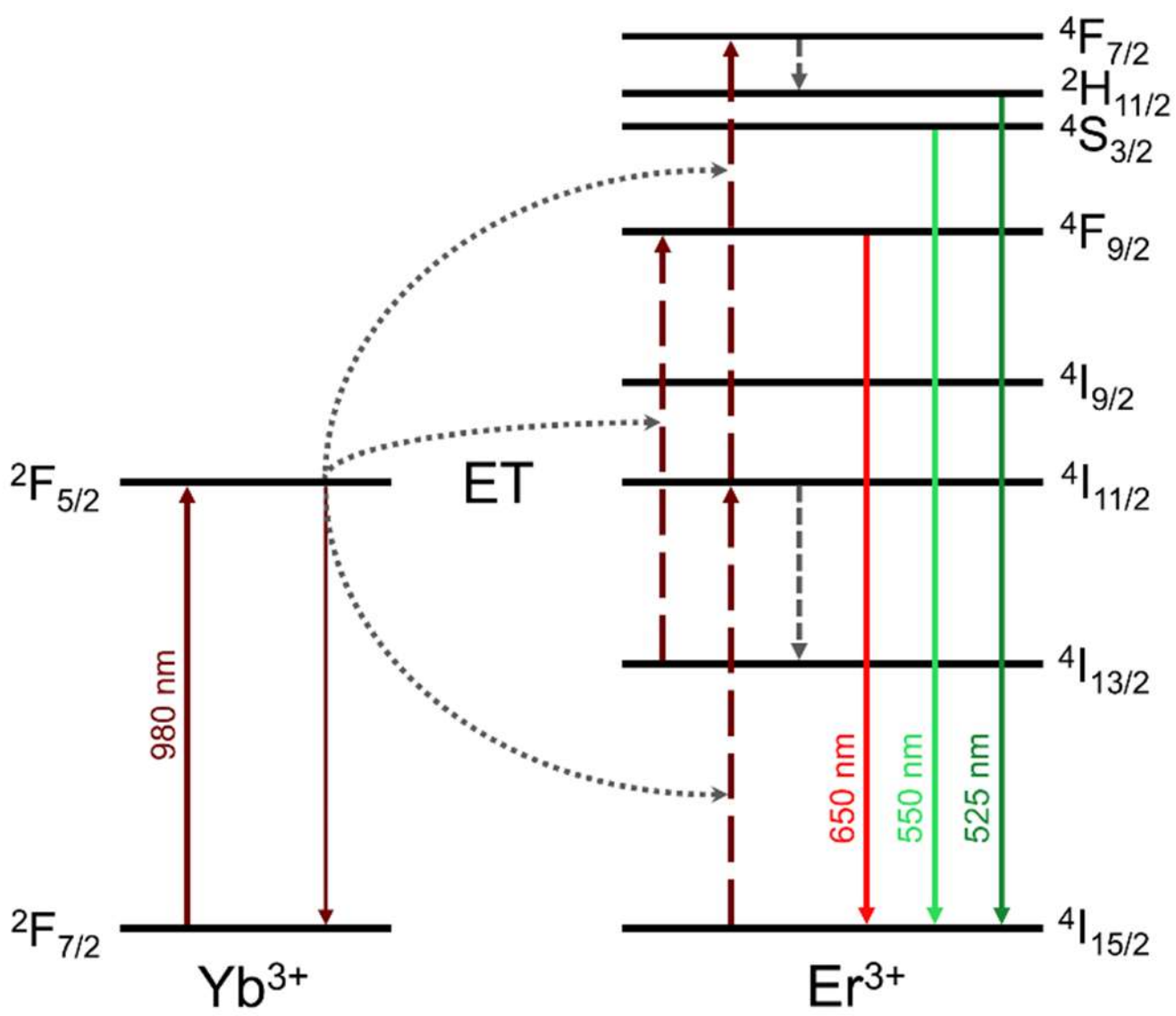

*Fig. S2. Schematic energy level diagram of the $Er^{3+}$/$Yb^{3+}$ co-doped system illustrating the up-conversion excitation and emission pathways under 980 nm excitation. Energy transfer (ET) from the $Yb^{3+}$ sensitizer to the $Er^{3+}$ activator populates higher-lying $Er^{3+}$ excited states, leading to visible up-conversion emission. Solid arrows indicate radiative transitions, dashed arrows denote excited-state absorption processes, and dotted arrows represent energy transfer between $Er^{3+}$ and $Yb^{3+}$ ions. The dominant $Er^{3+}$ emission bands at ca. 540 nm (525 and 550 nm), and 650 nm are highlighted.*

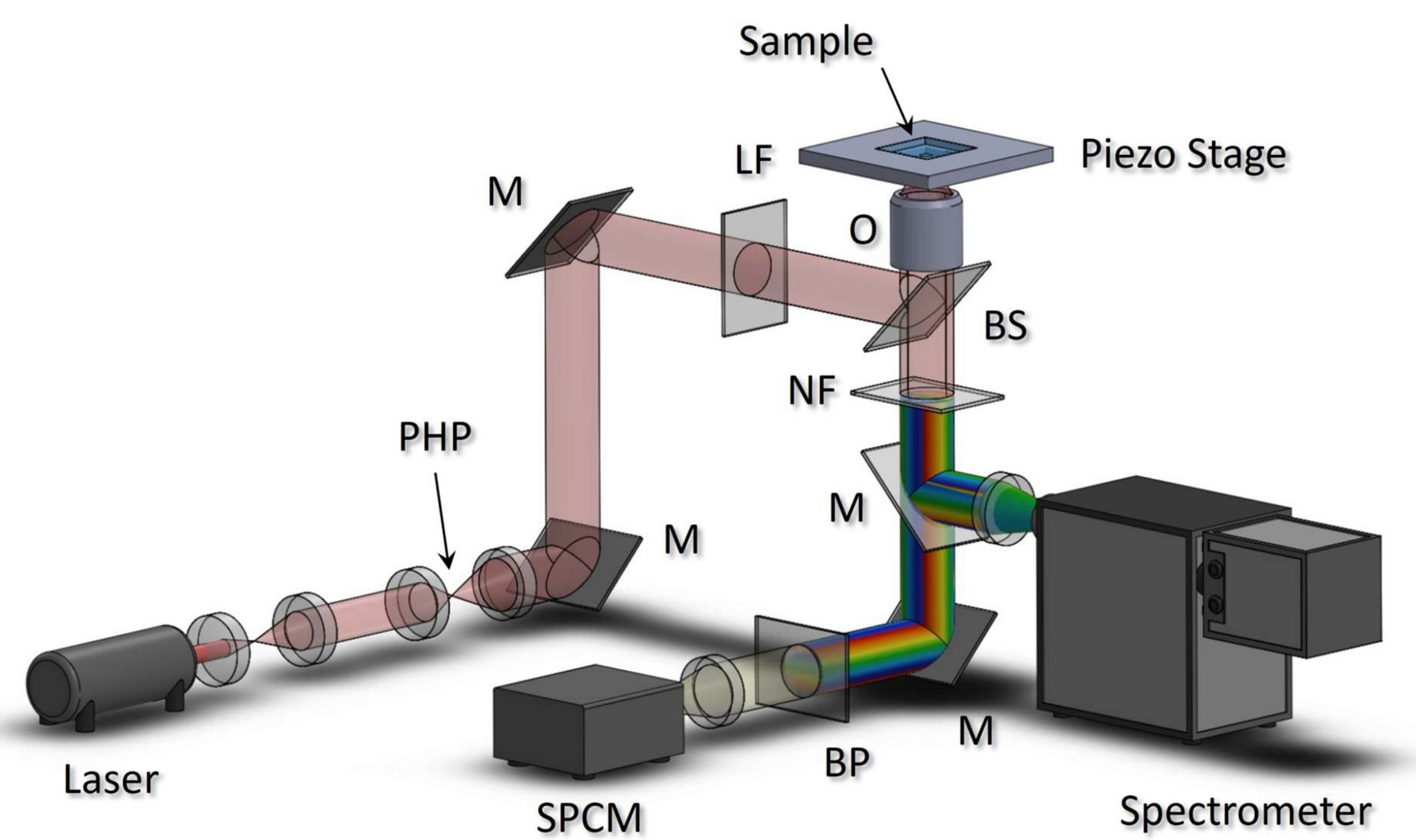

*Fig. S3. Schematic of the experimental setup. O – objective, BS – beam splitter, NF – notch filter, M – mirror, BF – band-pass filter, LF – laser filter, PHP – pinhole position. The excitation wavelength was $\lambda_{exc}$=980 nm, while detection was performed at $\lambda_{det}$=540/40 nm or $\lambda_{det}$=650/30 nm.*

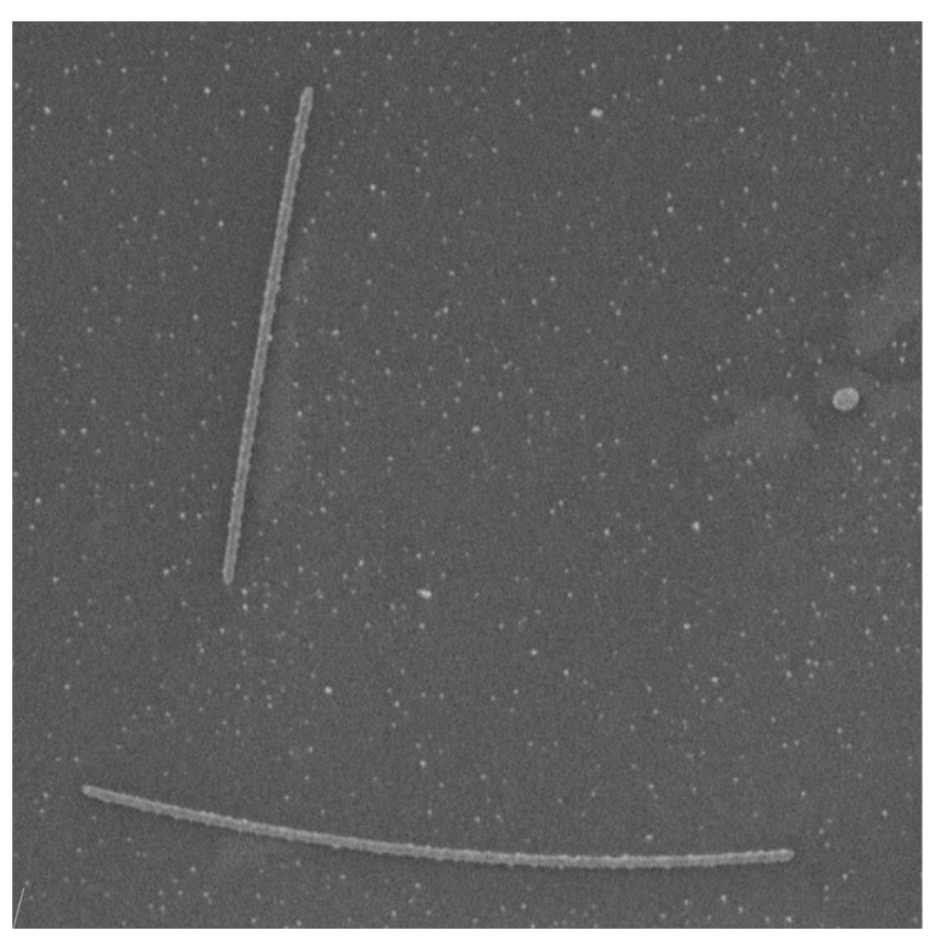

*Fig. S4. SEM image of sample S20 corresponding to Fig. 4b,e.*